\documentclass[preprint]{revtex4}

\usepackage{graphicx}%
\usepackage{dcolumn}
\usepackage{amsmath}

\begin{document}




\title{Modern calorimetry: going beyond tradition\\}

\author{Yoon Hee JEONG}
\affiliation{Department of Physics and electron
Spin Science Center, Pohang University of Science and Technology,\\
Pohang, Kyungbuk 790-784, S.  Korea\\}


\begin{abstract}
Calorimetry has been a traditional tool for obtaining invaluable
thermodynamic information of matter, the free energy. We describe
recent efforts to go beyond this traditional calorimetry: After
introducing dynamic heat capacity, we present the various
experimental methods to measure it. Applications and future
prospects are also given.
\end{abstract}

\maketitle

\section{INTRODUCTION}

At the turn of the 20th century  studies of heat capacity of
solids at low temperatures played an important role in revealing
the quantum character of nature \cite{einstein}. This example
vividly illustrates the power of calorimetry in science. As a
matter of fact calorimetric measurements reveal so great a deal of
information about matter that calorimetry, whose name originates
from the now obsolete caloric theory of heat described, for
example, by Lavoisier \cite{lavoisier}, has become an
indispensable tool for modern day research in chemistry, physics,
materials science, and biology \cite{wunderlich}. This is due to
the fact that it allows one to directly monitor the free energy
change of a given system as the external parameter such as
temperature varies. This unique ability of calorimetry follows
from the fact that it measures heat capacity as well as heat
itself.  Heat capacity at constant pressure is defined
thermodynamically by
\begin{equation}
C_p\,=\,\frac{dH}{dT}\,=\,T\frac{dS}{dT}
\label{eq:def}
\end{equation} where $T,\ H$, and $S$ are the temperature, the enthalpy,
and the entropy of a given system, respectively.  Thus, from the
measured heat (such as latent heat, heat of reaction, etc.) and
heat capacity, one may be able to reconstruct the free energy of
the system as a function of temperature or some other
thermodynamic variable.

Despite these positive attributes and historic roles of
calorimetry, one may naturally ask a question in the beginning of
a new millennium: is there any new scientific aspect in
calorimetry which will continue to attract attention and require
genuine efforts of researchers in the 21st century? This question
appears particularly pertinent if we consider the fact that
calorimetry is generally known as an  old discipline dealing with
thermodynamics of bulk materials, while a large portion of the
recent research efforts in various fields seem to be directed to a
understanding of the dynamic and/or local properties of
matter~\cite{lahnwitz2000}. In this short paper, I wish to address
the question from a personal point of view.

Measurement of a thermodynamic quantity is the usual notion that
is tied to calorimetry, since heat capacity is normally recognized
as a static thermodynamic quantity as defined above. However, it
is possible to go beyond this traditional understanding and
generalize heat capacity as a dynamic quantity. Indeed, it is
precisely my point that calorimetry should be generalized into the
dynamic and/or local regime to be able to deal with new scientific
issues in the coming years. In particular, the dynamic
generalization should empower calorimetry to deal with dynamic
phenomena in condensed matter, and at the same time may even
enable it in certain cases to become a local probe. Here, I wish
to describe the general principle and practice of {\it dynamic
calorimetry} (formally defined below), and only briefly touch upon
the generalization into the local regime. The details of local
calorimetry will be published elsewhere~\cite{local}.

Let us start with the formal definition of generalized {\it
dynamic heat capacity}. Of course, the term implies that heat
capacity is a quantity which may depend on either measuring time
or frequency. The concept of dynamic heat capacity appears natural
if one recalls that static thermodynamic quantities are
time-averaged (or ensemble-averaged).  In other words, they are
static not because they do not change in time, but because they
change too rapidly on the experimental time scale.  Suppose, for
instance, that a system contains a dynamic process relaxing with a
characteristic time $\tau$ which lies within our experimental time
window, then this will result in a time-dependent heat capacity
$C_{p}(t)$ depending on the time scale of measuring.  In the
frequency domain, this would lead to the complex
frequency-dependent heat capacity $C_p(\omega)$, and $C_p(\omega)$
of this system would show a dispersion in its real part and a peak
in its imaginary part at $\omega\,\approx\,\tau$.

One can utilize the linear response theory to formally define
dynamic heat capacity~\cite{jeong95}, and $C_p(\omega)$ can be
written in terms of enthalpy correlation function:
\begin{equation}
C_p(\omega)\,=\,C_p^0-{{i\omega}\over{k_BT^2}}\int_{0}^{\infty}dt
{\rm e}^{-i\omega t}<\delta H_R(0) \delta H_R(t)> \label{eq:dyn
def} \end{equation} where $C_p^0$ is the usual static heat
capacity, $\delta H_{R}$ is the slowly relaxing portion of
enthalpy fluctuation, and $<\,>$ denotes thermal averaging.  The
static heat capacity can be written as a sum of the fast part
$C_p^\infty$ (due to phonons in condensed matter which is our
concern), and the slow part: \begin{equation}
C_p^{0}\,=\,C_p^\infty\,+\, \frac{<\delta H_R^{2}>}{k_B T^2}\,.
\end{equation} From Eq.~\ref{eq:dyn def} it is easy to see that the slow
relaxation of enthalpy governed by the system dynamics is the
origin of the dynamic, or frequency-dependent, heat capacity and
therefore one can probe the slow dynamics of the system by
measuring the dynamic heat capacity.  Of course, it goes without
saying that this would be possible only if the slow part of
enthalpy fluctuation lies within the experimental time scale.
Throughout this paper, I shall use dynamic calorimetry to
collectively represent the calorimetric methods which allow the
measurements of dynamic heat capacity.

\section{GENERAL PRINCIPLE OF DYNAMIC CALORIMETRY}

In this section, I briefly expound the general principle of
dynamic calorimetry.  As usual with any dynamic experiment,
dynamic calorimetry can be conducted in either time or frequency
domain. The results obtained in the time domain are related to
those acquired in the frequency domain by the Fourier transform,
and vice versa.  It should also be pointed out that there are
again two ways, designated as type A and B, of measuring the
dynamic heat capacity of a given system in the time domain.  (And
the same is true in the frequency domain.) In type A, one may
attempt to induce a temperature jump in the system by thermally
attaching the system to a heat reservoir whose temperature is not
at the same value as that of the system, and measure the response
$\delta <H>$.  The variation of the enthalpy of the system can be
monitored by watching the heat flow ${\dot Q_{fl}}$ from the
reservoir to the system, since $\delta <H>\,=\,\int dt\, {\dot
Q_{fl}}$ by the first law of thermodynamics.  This is the method
used in commercial differential scanning calorimeters (DSC) where
temperature is varied according to a programmed schedule and the
resulting power is measured. The converse situation occurs in the
type B: One supplies a known amount of heat to the system and
monitors the ensuing temperature response with a sensor.

To expose the difference of the two methods more clearly, a
schematic diagram depicting the situation is shown in Fig.~1.
Fig.~1(a) illustrates the type A, where the system is in intimate
contact with the reservoir.  Now suppose that there is a step-like
temperature jump $\delta T$ in the reservoir.  The same $\delta T$
shift will be established in the system temperature via the heat
flow from the reservoir to the system.  If one carefully analyzes
${\dot Q_{fl}}$ as a function of time, one will find that it
consists of two parts: the instantly responding part, ${\dot
Q_{fl}^{\infty}}$, and the more slowly responding part, ${\dot
Q_{fl}^{R}}$.  In other words, the system takes up
$Q_{fl}^{\infty}\,=\,\int dt\, {\dot Q_{fl}^{\infty}} $, which is
due to the fast degrees of freedom (phonons in most cases of
interest), and brings up its temperature (or more accurately its
phonon temperature) by $\delta T$ instantly.  However, it takes
time for its slow degrees of freedom to respond and they will
absorb ${\dot Q_{fl}^{R}}$ gradually from the reservoir to reach
equilibrium~\cite{slow}. The opposite situation is encountered in
the type B, depicted in Fig.~1(b), where a constant amount of heat
is supplied to the system and the temperature of the system is
followed.  Here at the instant when heat is supplied, the energy
is shared only among the fast degrees of freedom, and then part of
energy slowly diffuses into the slow degrees of freedom.  Thus,
the temperature of the system first shows a step-like jump which
is then followed by a gradual decrease as a part of the supplied
heat energy is acquired by the slow degrees of freedom.

At this point it is worthwhile to emphasize that what one measures
with a sensor and calls the temperature of a system is actually
that of the fast degrees of freedom of the system, with which the
sensor is in direct thermal contact.  In what follows, I shall
enumerate various experimental methods which are currently in use
to measure dynamic heat capacity. Technical details of these
methods, however, are completely left out, and the interested
reader should refer to the literature~\cite{jeong97}.

\section{MODULATION CALORIMETRY}

\subsection{Traditional ac Calorimetry}

Traditional ac calorimetry~\cite{sullivan,hatta}, which belongs to
type B, is a calorimetric technique in which a small amount of
oscillating heat is supplied to a sample either by Joule heating
or heating by light and in either case the ensuing temperature
oscillation is measured. Since this temperature oscillation is
inversely proportional to the heat capacity of the sample in the
proper regime, one can determine the heat capacity of the sample
by carrying out measurements at a judicially chosen frequency.
While this method was mainly used to determine static heat
capacity of materials, it is natural to extend the technique as a
dynamic probe.

Consider a situation where an oscillating power of frequency
$\omega$ is applied to a sample of typical thickness $d$ and
thermal diffusivity $D$.  Then the thermal wavelength
$\lambda\,=\,\sqrt{D/\omega}$ characterize the length over which
the temperature oscillation decays, and the condition,
$d\,<<\,\lambda$, must be satisfied in order that the sample may
be regarded as oscillating in temperature as a whole, i.e.,
without a temperature gradient.  Since the lack of a temperature
gradient in the sample is prerequisite for an ac calorimeter to
work as a dynamic calorimeter, one may add metallic wires for a
liquid sample to enhance thermal conduction in the sample and may
make it as thin as possible for a solid sample. However, there
exists an unavoidable limit in this approach due to the fact that
the thermal wavelength decreases as frequency increases, and
eventually beomes shorter than $d$.  Since the geometry of a
sample here is not particularly well defined, it is very hard to
deal with the situation where a temperature gradient exists in the
sample.  So, the method is usually limited to less than 1 Hz for
liquids or dielectric crystals and perhaps tens of Hz for metallic
solids.

\subsection{3$\omega$ Calorimetry}

In order to extend the dynamic range of calorimetry, one has to
deal with the temperature gradient effect, rather than try to
avoid it. One method, which has been successful, is to take
advantage of a well-defined geometry with thickness of a sample
much larger than the thermal wavelength.  Here one faces a
heat-diffusion problem in a semi-infinite medium with a heater on
the surface that generates the oscillating power.  The
well-defined geometry of the sample in this case enables one to
solve the heat diffusion equation exactly, and it turns out that
the surface temperature oscillation is determined by the thermal
properties of the medium. Thus, one can extract the dynamic heat
capacity by measuring the surface temperature oscillation as a
function of the frequency of the oscillating power.

The 3$\omega$ method adopts a clever idea of using a metallic
heater on the surface simultaneously as a temperature sensor that
measures the surface temperature
oscillation~\cite{birge,jung,moon1}. Since the sensor is at the
same position as the heater (it is the heater itself), one does
not encounter a problem associated with the fact that the
temperature oscillation decays within the thermal wavelength which
becomes short at high frequencies. Thus, one can raise the upper
limit of the dynamic range of modulation calorimetry.  The reason
why it is called the 3$\omega$ method is the following: since one
uses Joule-heating by passing an electric current at a certain
frequency $\omega$ through a metallic heater on the surface of a
material, the surface temperature oscillation occurs at 2$\omega$.
This temperature oscillation induces resistance oscillations in
the heater at the same frequency, and then the voltage drop across
the heater itself would include a term at 3$\omega$.  This
3$\omega$ signal is detected to yield the dynamic heat capacity of
the material at 2$\omega$. The currently available frequency in
3$\omega$ calorimetry ranges approximately  from 0.01 Hz to 50
kHz~\cite{jung2}.

\subsection{Differential Scanning Temperature Modulated Calorimetry}

A differential scanning temperature modulated calorimeter (DSTMC)
is a commercially available calorimeter, which allows one to do
dynamic calorimetry~\cite{nomen}.  Differential scanning
calorimeters are distinct from the modulation calorimeters
described above in that they belong to type A.  A DSC measures the
heat flow to a sample, which is necessary to keep up with the
prescribed temperature variation, normally linear heating or
cooling, of the reservoir (or the block). In DSTMC, this
conventional temperature program has a superimposed periodic
temperature variation, so that \begin{equation}
T(t)\,=\,T_{0}\,+\,\beta t\,+\,A\sin\omega t \end{equation} where
$T_{0}$ is the initial temperature, $\beta$ is the underlying
heating rate, and $A$ and $\omega$ are the amplitude and frequency
of modulation.  By measuring the response of the sample to this
temperature variation one can, at least in principle, obtain
dynamic information as a function of frequency.  Although DSTMC
has contributed to the widespread use of modulation calorimetry,
there remains some problems to be clarified before it can be used
reliably as a dynamic calorimetric tool.  A particualrly serious
problem is that the signal one gets from DSTMC seems to be not
only due to the sample properties, but also due to the calorimeter
itself.  Currently much effort is being spent on this
problem~\cite{lahnwitz}.

\subsection{Peltier Calorimetry}
Peltier calorimetry is an ac calorimetric method developed
specifically for the heat capacity measurements, dynamic as well
as static, of minute samples~\cite{moon00}. Most conventional
calorimetric methods require indispensable addenda (heater and
sensor) to be put on a sample and the mass of these addenda may
even be greater than that of the sample in case of minute sample
masses. The results of this situation are the loss of sensitivity
and accuracy in determining the absolute value of heat capacity.
In other words, the unavoidable calibration procedure (removing
the background due to the addenda)  introduces uncertainty in both
accuracy and precision. Peltier calorimetry overcomes this
difficulty by utilizing the well-known Peltier effect of a
thermocouple as an ac power source; the junction of a very thin
thermocouple (of diameter 1 mil or less) attached to a sample is
so tiny that its mass is totally negligible and therefore it can
play the role of an ideal heater. It should be pointed out that a
thermocouple can not only heat, but also cool the system, whereas
other power sources such as Joule heating or heating by light can
only heat the system. The temperature oscillation is sensed with
another thermocouple which again does not disturb the system.
Peltier calorimetry is superior to other modulation methods in
that the experimental setup is simple, the average temperature of
a sample is at, not above as in traditional ac calorimetry, the
bath temperature, and it directly yields absolute values of the
heat capacity of sub-milligram samples without the necessity of a
calibration procedure.

A novel extension of Peltier calorimetry is the Peltier thermal
microscope, which would enable one to measure local thermophysical
properties of matter at small length scales.   For this purpose,
we have shown that a single junction can be used  simultaneously
as both power source and sensor~\cite{local}. It must be stressed
that this represents a conceptual deviation from the normal notion
that a power source and a sensor are required separately in
calorimetry and constitutes a major step toward a local
calorimeter. Then, substitution of a thermocouple for a tip in an
atomic force microscope would constitute a thermal
microscope~\cite{majumdar}. A potentially  very useful
characteristic of the Peltier microscope is that one can change
the probing length scale (thermal diffusion length) by changing
the frequency of a current in the thermocouple. The successful
development of the Peltier thermal microscope would be an exciting
and important event for this so-called nano-age where the
sub-micrometer local thermophysical properties are in great
demand.

\section{ADIABATIC CALORIMETRY}
\subsection{Time Domain Dynamic Calorimetry}

One of the most difficult tasks in calorimetry is to contain the
leakage of heat, since there does not exist a perfect thermal
insulator. Even if there were one, radiation would still have to
be contained. In adiabatic calorimetry, heat leak from a sample to
the surroundings is controlled to zero as tightly as possible.
Since Nernst first constructed an adiabatic
calorimeter~\cite{nernst}, adiabatic calorimetry has been the most
accurate method of determining static heat capacity.  In addition
to this capability an adiabatic calorimeter can be used as a means
of dynamic calorimetry in time domain~\cite{wunderlich57}.  In
time domain dynamic calorimetry, a certain amount of heat is
applied instantaneously to a sample and its temperature response
is examined in real time. Recall the situation depicted in
Fig.~1(b). Since the adiabaticity is maintained all the time, the
variation of temperature in time is nothing but time dependent
heat capacity. By varying the values of applied energy, one can
induce temperature jumps of varying size. While it is not easy to
apply a large amount of power to a sample in modulation
calorimetry, there is no difficulty in varying the temperature
jump sizes in time domain dynamic calorimetry.  Thus this feature
of the time domain method allows one to study nonlinear
temperature relaxation~\cite{moon97}.

\subsection{Scanning Adiabatic Calorimetry}

One classical method to extract dynamic calorimetric information
from a system is to perform scanning experiments. In this method
the experimental time scale is set by the scanning rate $dT/dt$.
Traditionally commercial differential scanning calorimeters have
been utilized for this purpose~\cite{moynihan}, but  adiabatic
calorimeters can also be operated in the scanning mode.  Since the
adiabatic condition between the sample and the shield must be
maintained during operation, it is usual that the scanning mode is
adopted in a heating experiment~\cite{thoen}. However, it is
necessary to measure heat capacity during cooling as well as
heating at various scanning rates to extract dynamic information
if a substance shows frequency dependence in heat capacity.  This
situation motivated us to develop a method to use an adiabatic
calorimeter in a rate-scanning mode at variable rates  during both
heating and cooling~\cite{moon2}.  In the cooling mode, a constant
temperature difference is maintained, and then there is a heat
flow from the sample to the shield.  It turns out that one can
control the cooling rate by varying the size of the temperature
difference. The heat flow in this case is not a leak, but is used
as a negative input power in a controlled fashion.  The
calorimeter used in this way can be operated in both heating and
cooling modes in the scanning range of 0.01 - 2 K/min.  It is
noted that adiabatic dynamic calorimetry belongs to type B.

\section{APPLICATIONS AND FUTURE OUTLOOK}

In this paper I have outlined dynamic calorimetric techniques
currently in use. Modulation calorimetry of type B, where one
supplies a small oscillating heat input to a sample, allows one to
measure the dynamic heat capacity in addition to the usual static
heat capacity under equilibrium conditions.  One can measure
dynamic heat capacity directly with a traditional ac calorimeter.
However, the dynamic range provided by this method is quite
limited, below 1 Hz in many cases with solids as well as liquids.
To extend the dynamic range, one can use the 3$\omega$ method.
While this method raises the upper frequency limit to tens of kHz,
one has to pay the price of being involved with thermal
conductivity.  To get dynamic heat capacity data, one has to
perform separate measurements of the thermal conductivity.  A
commercial DSTMC, which belongs to type A, is a convenient tool
for qualitative purposes, but it seems that there remains some
problems to be solved before it can become a quantitative dynamic
calorimeter. A Peltier calorimeter is an attractive method for
obtaining the dynamic heat capacity of minute samples. Currently,
a Peltier thermal microscope, a variation of the Peltier
calorimeter, is under development. An adiabatic calorimeter is the
most accurate tool for heat capacity and latent heat measurements.
In particular, the latent heat information is very difficult to
obtain with other methods. With the advance in computer control
and electronics technology, the operation of adiabatic
calorimeters has become less tedious and less time-consuming. This
advance has allowed the time domain dynamic calorimetry where one
follows the temperature change of the system under adiabatic
conditions, after a certain amount of heat is given to a system.
Enthalpy relaxation, including nonlinear relaxation, can be
studied in real time with adiabatic calorimeters.

As for the future research, one may try to extend the dynamic
range of the 3$\omega$ method above the current frequency limit,
tens of kHz. If successful, then the thermal wavelength can enter
the submicrometer regime in dielectric materials. This would allow
the characterization of thermal properties of thin films with
submicrometer thicknesses. This kind of information is not only
scientifically interesting (heat capacity and thermal conductivity
of a sample with restricted geometry) but also technologically
important (efficient heat transport in thin films). The
development of noncontact heating at higher frequencies than is
available with the present light chopping would be very helpful in
enhancing the applicability of modulation calorimeters. Various
techniques for measuring thermophysical properties of thin films
utilizing advanced laser technology seem very
promising~\cite{baba}. Although we have generalized heat capacity
temporally by defining a dynamic heat capacity, one may also
attempt to generalize heat capacity spatially by defining a heat
capacity which is wavenumber as well as frequency dependent. This
full generalization of heat capacity would constitute a challenge
theoretically as well as experimentally. Development of local
calorimeters with nanoscale or at least sub-micrometer resolution
would be an exciting event. This is particularly urgent to keep up
with the current technology trend heading toward the nanometer
regime.

Even with the current dynamic range, it is recognized that dynamic
calorimetry has wide applications in diverse fields. For example,
biological systems, where slow dynamics is often an important
occurrence, may be suitable to be studied by dynamic calorimetry.
Also of keen interest and a good candidate for dynamic calorimetry
is the melting phenomenon, where one can expect contributions from
time-dependent processes. Here the latent heat effect and specific
heat effects are usually entangled in calorimetric measurements to
make the interpretation of data obscure; however, by combining
various methods presented here one can hope to disentangle them.
One can also expect to find use of dynamic calorimeters in
situations involving solids: global thermal hysteresis phenomenon
in incommensurate ferroelectrics, heat conduction in porous media
and so on.

\begin{acknowledgments}
This work was supported by KOSEF (1999) and the BK21 program of
POSTECH.
\end{acknowledgments}

\begin{figure}
\includegraphics[width=10cm]{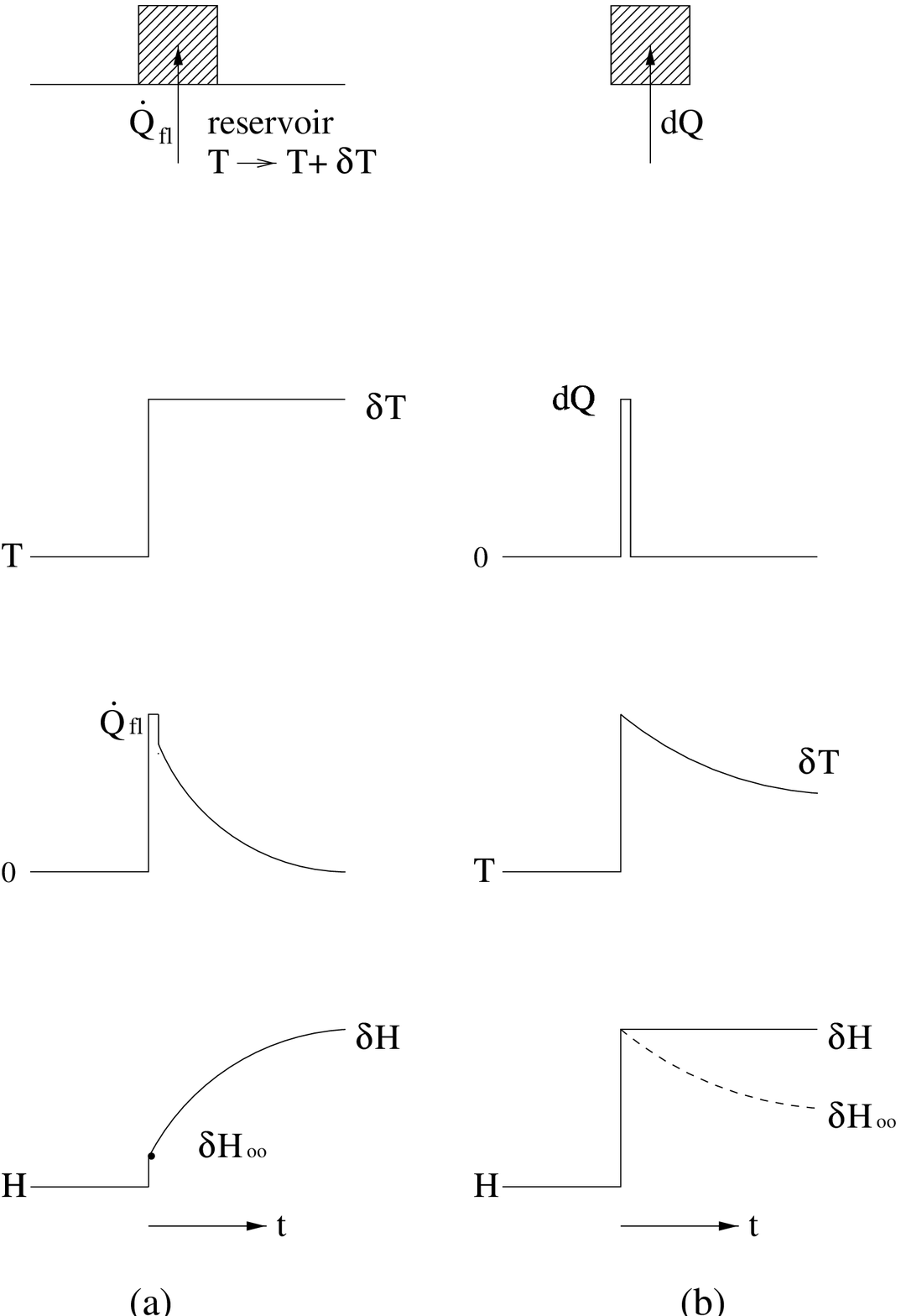}
\caption{Schematic diagram showing two methods of measuring
dynamic heat capacity. (a) Type A: A system is thermally attached
to the reservoir and the temperature of the reservoir jumps from
$T$ to $T+\delta T$ at a certain instant (top and second figure).
The heat flow ${\dot Q_{fl}}$ from the reservoir to the system, as
described in the third figure, is measured.  An instantaneous heat
absorption, corresponding to $\delta H_\infty$, by the fast
degrees of freedom is followed by the gradual heat flow caused by
the absorption due to the slow degrees of freedom (bottom figure).
(b) Type B: A certain amount of heat, $dQ$, is given to a system
(as sketched in the top and second figure), and the change in $T$
is followed in time.  $T$ jumps instantaneously, and this jump is
accompanied by a gradual decrease as the supplied heat energy is
shared by the slow degrees of freedom (third figure). The decrease
in $\delta H_\infty$ appears as an increase in the internal
enthalpy $\delta <H_R>$ by the same amount (bottom figure).}
\label{fig1}
\end{figure}


\begin{references}
\bibitem{lahnwitz2000} A consensus regarding the current status of calorimetry
was reached by the participants of
the 6th L\"ahnwitz seminar on calorimetry in the final discussion
session.
\bibitem{slow} Here we are arbitrarily dividing degrees of freedom
in condensed matter into fast and slow ones. Generally, the phonon
time scale (of order of pico-second) would be instantaneous for
the present purpose, but the slow time scale may range from
hundreds of seconds (mHz) to micro-second (MHz).
\end{references}
\end{document}